\documentclass[aps,prb,showpacs,revsymb,amsfonts,amssymb,twocolumn,
groupaddress,footinbib,floatfix]{revtex4-1}

\usepackage{graphicx}
\usepackage{amsfonts}
\usepackage{amssymb}
\usepackage{amsmath}
\usepackage{verbatim}
\usepackage{yfonts}

\usepackage[pdftex,bookmarks,colorlinks,breaklinks]{hyperref}  
\hypersetup{linkcolor=blue,citecolor=red,filecolor=green,urlcolor=green} 

\begin{document}


\title{Campbell penetration in the critical state of type II superconductors}

\author{R.\ Willa}
\author{V.B.\ Geshkenbein}
%
%
\author{G.\ Blatter}
\affiliation{Institute for Theoretical Physics, ETH Zurich, 
8093 Zurich, Switzerland} 
%

\date{\today}

\begin{abstract}
The penetration of an $ac$ magnetic signal into a type II superconductor
residing in the Shubnikov phase depends on the pinning properties of Abrikosov
vortices. Within a phenomenological theory, the so-called Campbell penetration
depth $\lambda_{\rm \scriptscriptstyle C}$ is determined by the curvature
$\alpha$ at the bottom of the effective pinning potential. Preparing the
sample into a critical state, this curvature vanishes and the Campbell
length formally diverges. We make use of the microscopic expression for the
pinning force density derived within strong pinning theory and show how flux
penetration on top of a critical state proceeds in a regular way.
\end{abstract}


\maketitle


\section{Introduction}\label{sec:intro}

Transport \cite{onnes_11} and magnetization \cite{meissner_33} measurements
are well known tools for the basic phenomenological characterization of
superconductors.  The diamagnetic screening in a superconductor involves
multiple aspects: a weak magnetic field penetrates to the material over the
length $\lambda_{\rm\scriptscriptstyle L}$, the (London) penetration depth
\cite{london_35}, which provides access to the superfluid density $\rho_s$ and
is typically of sub-micrometer size.  In type II superconductors, a magnetic
field $H$ penetrates the material through quantized flux-lines or Abrikosov
vortices \cite{abrikosov_57} which arrange in a triangular lattice defining
the Shubnikov phase \cite{shubnikov_37}. Testing this phase via a small
$ac$-magnetic field produces a normal response described by the skin-effect,
with a reduced (flux-flow) resistivity $\rho_\mathrm{ff}$ entering the usual
expression for the skin-depth $\delta(\omega)$. In real materials, vortices
get pinned by material defects, thereby establishing the desired critical
current density $j_c$ below which vortices are trapped.  In this situation, an
external $ac$-field probes the pinning landscape (or pinscape), as the latter
now determines the field penetration over the scale $\lambda_{\mathrm{
\scriptscriptstyle C}}$ (typically $1$ -- $100$ micrometers), the Campbell
length\cite{campbell_69}.  The linear response in the Campbell regime assumes
that the vortex displacements induced by the $ac$-magnetic field $h_{ac}$ are
much smaller than the characteristic pinning length such that the Campbell
response probes the pinning wells. This condition implies that the currents
$\delta j \sim c h_{ac}/\lambda_{\mathrm{ \scriptscriptstyle C}}$ induced by
the vortex displacements are much smaller than the system's critical current
$j_{c}$, $\delta j \ll j_c$. Increasing the field strength $h_{ac}$ to large
values capable of changing the direction of the critical state periodically,
i.e., $\delta j > j_{c}$, the response is described by the Bean
model\cite{bean_62}. In this limit, the Bean penetration depth
$\ell_{\mathrm{\scriptscriptstyle B}} \sim c h_{ac}/4\pi j_c$ depends on the
field amplitude $h_{ac}$ and hence generates a higher harmonic signal in the
magnetic response.

\begin{figure}[t!]
\includegraphics[width=0.47\textwidth]{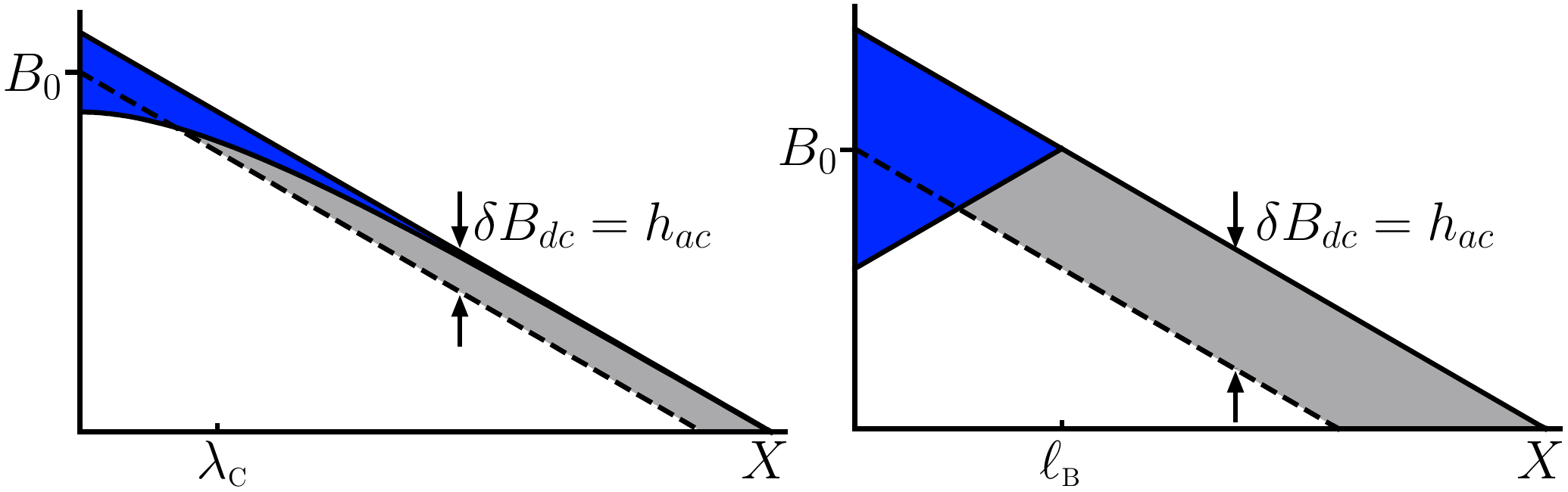}
\caption{
Sketch of the linear Campbell (left panel) vs. non-linear Bean (right panel)
response of a vortex critical state subject to an $ac$ magnetic field
$h_{ac}$. The gray-shaded areas show the change of the $dc$ field deep inside
the sample after the initial field penetration at short times. The blue
regions indicate the magnetic field oscillations at large times penetrating
the sample to the depths $\lambda_{\mathrm{\scriptscriptstyle C}}$ and
$\ell_{\mathrm{\scriptscriptstyle B}}$, respectively. The cross-over between
the Campbell- and Bean regimes with increasing field amplitude $h_{ac}$ occurs
when the induced current $\delta j$ near the sample surface is of the order of
$j_{c}$, i.e., when $h_{ac} \gtrsim j_{c} \lambda_{\mathrm{\scriptscriptstyle
C}} / c$.}
\label{fig:CvsB}
\end{figure}

Within a phenomenological model of the Campbell response, the dynamics of
vortices in a random potential landscape is reduced to the motion of a vortex
in an effective defect potential. Vortices probe the bottom of the pinning
potential and the Campbell penetration depth $\lambda_{\rm \scriptscriptstyle
C} \propto 1/\sqrt{\alpha}$ involves the curvature $\alpha$ at the minimum of
the effective pinning well \cite{campbell_69}. This description becomes quite
problematic when dealing with the technologically most relevant vortex
configuration, the critical state \cite{bean_62} realizing the maximally
possible current flow (the critical current density $j_c$) before depinning.
Indeed, within this approach, the critical state is characterized by a
vanishing\cite{prozorov_03} curvature $\alpha(j) \propto (j_{c} - j)^{1/2} \to
0$ and thus implies a diverging Campbell length $\lambda_{\rm
\scriptscriptstyle C}$.  This result is not satisfactory in two respects:
first, the predicted divergence is not observed in experiments
\cite{prozorov_03}; an explanation that flux creep prevents probing of the
proper critical state is hardly applicable to the early work by
Campbell\cite{campbell_69} on low-$T_{c}$ Pb-Bi alloys. Second, it turns out
that the Campbell length in the critical state can be even smaller than that
in the field-cooled state\cite{prozorov_13,willa_15b}.

In this paper, we make use of a microscopic theory in order to reconcile the
apparent divergence of the penetration depth $\lambda_{\rm\scriptscriptstyle
C} \propto 1/\sqrt{\alpha}$ in the critical state with a regular vortex
dynamics. To this end, we describe the $ac$-magnetic penetration within the
strong pinning framework \cite{willa_15b} and determine the dynamical evolution
of the vortex state when probing a critical state with a small-amplitude and
low-frequency $ac$-magnetic field $h_{ac} e^{-i\omega t}$. We show that in
this situation the $ac$-response involves a transient region where vortices
first penetrate throughout the sample in the form of diffusive flux pulses.
The penetrated flux produces an upward shift of the critical state profile;
once this $dc$ shift reaches the maximal field amplitude $h_{ac}$, see
Fig.~\ref{fig:CvsB} (left), the $ac$-field only lowers the magnetic field at
the sample edge and we arrive at a \textit{finite} $ac$ penetration depth
$\lambda_{\rm\scriptscriptstyle C} \propto 1/\sqrt{\Delta f_\mathrm{pin}}$,
with $\Delta f_\mathrm{pin}$ the jump in the microscopic pinning force for the
critical state as given by the strong pinning theory. While, qualitatively, a
similar picture describes the non-linear $ac$ response at large field
amplitudes $h_{ac} > j_{c} \lambda_{\rm\scriptscriptstyle C} / c$ as described
by the Bean model\cite{bean_62} and illustrated in the right panel of Fig.\
\ref{fig:CvsB}, the quantitative description and results are very different
for the Campbell and Bean penetration regimes.

In the following, we briefly review (Sec.\ \ref{sec:ac}) the $ac$-magnetic
response of a type II superconductor in the Shubnikov phase and its
microscopic extension using the result from strong pinning theory.  In Section
\ref{sec:indyn} we describe the transient regime of the $ac$-response with its
vortex penetration, the central topic of this paper.  Section \ref{sec:sc}
gives a short summary and conclusions.

\section{$ac$ Response and Campbell Length}\label{sec:ac}

An $ac$ dynamical field enters a homogeneous metallic sample over the
skin-depth $\delta (\omega) \approx \sqrt{c^2/2\pi \mu \omega \sigma}$, with
$\mu$ and $\sigma$ the materials' magnetic permeability and conductivity,
respectively. For the inhomogeneous state of a field-penetrated type II
superconductor, it is the behavior of vortices which determines the $ac$
response. In a real, i.e., defected material, the dynamics of pinned vortices
is dictated by the pinning potential opposing their motion; within a
phenomenological model, the force density $F_\mathrm{pin} \approx -\alpha x$
defines the penetration depth $\lambda_{\rm\scriptscriptstyle C}$ at small
frequencies $\omega$,
\begin{align}\label{eq:lC}
   \lambda_{\rm \scriptscriptstyle C}^{2} = \frac{B^2}{4\pi \alpha},
\end{align}
where $B$ is the mean penetrated field.  The above result for
$\lambda_{\rm\scriptscriptstyle C}$ was first given by Archie Campbell
\cite{campbell_69}, together with supporting experimental data on the
$ac$-magnetic response of a superconductor.

For a brief derivation of the result (\ref{eq:lC}), we consider a
superconductor occupying the half-space $X>0$ in the presence of a magnetic
field involving both $dc$ and (a small) $ac$ component $H(t) = H_0 + h_{ac}
\exp(-i\omega t)$ directed along $Z$ (for consistency with Ref.\
\onlinecite{willa_15b} we use capital-letter coordinates in describing the
macroscopic situation).  This field penetrates to the sample in the form of
vortices producing an average magnetic induction $B(X,t)$.  The screening
current density $j$ in the superconductor flows along the $Y$-axis and exerts
a Lorentz force density $F_{\rm \scriptscriptstyle L} = j B/c$ directed into
the sample.  In a stationary state, the Lorentz force density $F_{\rm
\scriptscriptstyle L}$ has to be balanced by the pinning force density
$F_\mathrm{pin}$, otherwise vortices move dissipatively. Introducing the
\emph{macroscopic} displacement field $U(X,t)$ of the vortex system, the force
balance equation takes the form
\begin{align}\label{eq:eom}
   \eta \partial_t U = F_{\rm \scriptscriptstyle L} (j,U) + F_\mathrm{pin} (X,U),
\end{align}
with $\eta$ denoting the viscosity \cite{bardeen_65}. The magnetic induction
and current can be split into a $dc$ part and a contribution from the external
$ac$ drive, $B(X,t) = B_0 +\delta B(X,t)$ and $j(X,t) = j_0 + \delta j(X,t)$,
where $B$ is driven at the boundary, $B(0,t) = B_0 + h_{ac} \exp(-i\omega t)$.
The $ac$-magnetic induction $\delta B$ relates to $\delta j$ via Amp\`ere's
law, $\partial_{\rm \scriptscriptstyle X} \delta B = -(c/4\pi) \delta j$, and
to the displacement $U$ via the change in vortex density, $\delta B/B_0
\approx -\partial_{\scriptscriptstyle X} U$. For a critical state, the $dc$
current density $j_0$ is maximal and hence equals the critical current density
$j_c$. The latter is compensated by the maximal (or critical) pinning force
density $\max F_\mathrm{pin} = - F_c$, with $F_c = j_c B/c$. Rewriting the Lorentz
force density through $U$ and denoting deviations from maximal pinning by
$\delta F_\mathrm{pin} \equiv F_c - F_\mathrm{pin}(U)$, we arrive at the dynamical
equation of the form,
\begin{align}\label{eq:eomU}
   \eta \partial_t U - (B_0^2/4\pi) \partial^{2}_{\scriptscriptstyle X} U 
   - \delta F_\mathrm{pin} (U) = 0,
\end{align}
Alternatively, this equation can be obtained starting from the equation of
motion of individual vortices. Averaging over many inter-vortex spacings
$a_{0}$, one then arrives at the above expression. With the chosen field and
current directions, the vortex-vortex interaction [second term in Eq.\
\eqref{eq:eomU}] only involves the bulk compression modulus
$c_{11}(\boldsymbol{k}=0) = B_{0}^{2}/4\pi$, while contributions of the shear
and tilt moduli are averaged to zero.

The non-trivial part in arriving at an explicit equation for $U(X,t)$ is the
functional dependence of the pinning force density $\delta F_\mathrm{pin}
(U)$. Assuming vortices trapped in a harmonic pinning potential, Campbell
\cite{campbell_69} introduced the phenomenological Ansatz $\delta
F_\mathrm{pin}(U) = - \alpha U$, with $\alpha$ describing the curvature of the
effective pinning potential. The resulting differential equation is of the
driven-diffusive type and easily solved,
\begin{align}\label{eq:UC}
   U(X,t) &=  \lambda_{\rm \scriptscriptstyle C}(h_{ac}/B_0) 
   e^{-X/\lambda_{\rm \scriptscriptstyle C}} e^{-i \omega t}
\end{align}
with
\begin{align}\label{eq:lCo}
   \lambda_{\rm \scriptscriptstyle C}^2(\omega) 
          &= \frac{B_0^2}{4\pi}\frac{1}{\alpha - i \omega \eta}.
\end{align}
At small frequencies, we obtain the Campbell length $\lambda_{\rm
\scriptscriptstyle C} = \lambda_{\rm \scriptscriptstyle C} (\omega = 0)$ as
given in Eq.\ (\ref{eq:lC}); at high frequencies we make use of the Bardeen
formula $\eta = BH_{c2}/c^2 \rho_n$ to arrive at the skin depth
$\delta(\omega) \approx (c^2/2\pi \omega \sigma_\mathrm{ff})^{1/2}$ with
$\sigma_\mathrm{ff} = \sigma_n H_{c2}/B$ the flux flow conductivity,
$\sigma_n$ the normal state conductivity, $H_{c2} = \Phi_0/2\pi\xi^2$ the
upper critical field, and $\xi$ the coherence length of the superconductor.

\begin{figure}[t!]
\includegraphics[width=7.5cm]{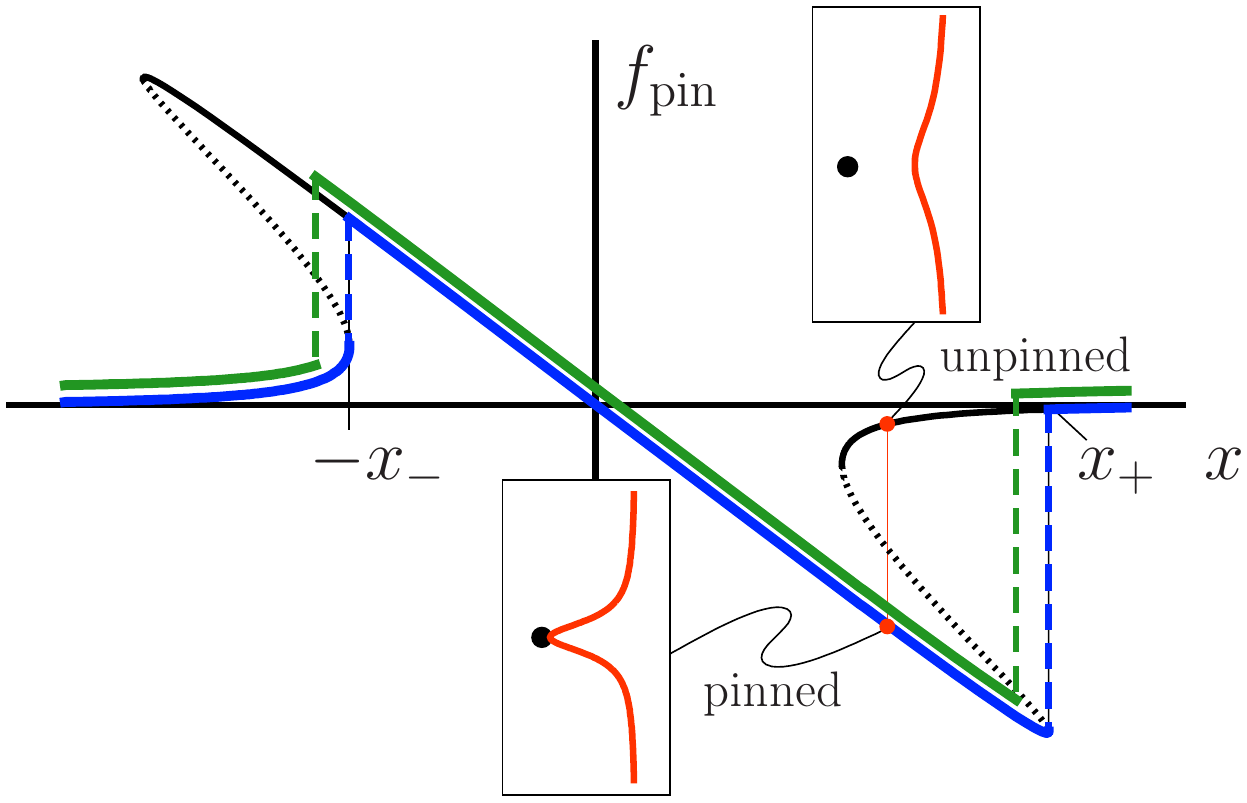}
\caption{Microscopic effective pinning force $f_{\mathrm{pin}}$ with bistable
pinned and unpinned solutions as a function of the pin-vortex distance $x$
(dotted lines denote unstable solutions). For randomly positioned defects and
the vortex lattice in the critical state, the occupation of the force branches
(blue) produces the maximal restoring force $F_{\mathrm{pin}} = n_{p}\langle
f_{\mathrm{pin}}\rangle = -F_{c}$. A macroscopic displacement $U > 0$ of all
vortices in the direction of the Lorentz force does not change the branch
occupation and $\delta F_{\mathrm{pin}} = 0$. A displacement $U<0$ against the
Lorentz force, however, modifies the branch occupation (green) and reduces the
pinning force $\delta F_{\mathrm{pin}} \propto \Delta f_{\mathrm{pin}} U$, see
Eq.~\eqref{eq:dFp}.}
\label{fig:fpin}
\end{figure}

Here, we go beyond this phenomenological theory and make use of the expression
for the restoring force $\delta F_\mathrm{pin}$ derived from strong pinning
theory \cite{willa_15b}.  The latter goes back to early work of Labusch
\cite{labusch_69} and of Larkin and Ovchinnikov \cite{larkin_79} and has
attracted quite some interest over the recent years
\cite{blatter_04,koshelev_11,thomann_12}. Within the framework of strong
pinning theory, vortices are pinned by individual defects, thus allowing for a
quantitative description of pinning related phenomena. The crucial feature
appearing within strong pinning is the strong deformation of vortices giving
rise to bistable solutions of the force equation balancing the elastic vortex
energy against the pinning energy due to the defect; the appearance of such
bistable solutions, quantitatively formulated in the Labusch
criterion\cite{labusch_69} $\kappa = 1$, then separates strong ($\kappa > 1$)
from weak ($\kappa < 1$) pinning (the Labusch parameter $\kappa \sim
f_p/\bar{C}\xi$ measures the relative strength of the pinning force $f_p$ of
one defect as compared to an effective elasticity $\bar{C}$ of the vortex
lattice).

Within strong pinning theory, one studies how a representative vortex embedded
in the vortex lattice gets locally deformed and pinned due to the presence of
a defect \cite{willa_15b}; the macroscopic pinning force density
$F_{\mathrm{pin}}$ then results from proper averaging of the microscopic
effective pinning force $f_{\mathrm{pin}}$. More specifically, when dragging a
vortex across an individual defect, the vortex jumps into the pinning
potential at $-x_{-}$, $x_{-} \sim \xi$, and remains pinned therein until the
deformation becomes too large and the vortex snaps out of the pin at $x_{+}
\sim \kappa \xi$, see Fig.~\ref{fig:fpin}.  Assuming a small density $n_p$ of
pinning centers, one can ignore interactions between defects and the pinning
force density $F_{\mathrm{pin}} = n_{p} \langle f_{\mathrm{pin}} \rangle$
derives from a simple average over the microscopic pinning states with
vortices occupying pinned and unpinned branches of the bistable pinning
landscape.  A vortex system in the (Bean) critical state is defined through
the maximal averaged pinning force density $F_c$ produced by the pins. In this
state, the vortex configuration is critical, i.e., when shifting the vortex
system in the direction of vortex penetration $U > 0$ there is no change in
pinning force as the latter is already maximal,
\begin{align}\label{}
   \delta F_\mathrm{pin} (U > 0) = 0.
\end{align}
On the other hand, moving vortices opposite to
the critical slope, i.e., for $U < 0$, the branch occupation rearranges, see
Fig.\ \ref{fig:fpin}, and the pinning force density is diminished by
\cite{willa_15b}
\begin{align}
   \delta F_\mathrm{pin}(U < 0) &\approx -n_p (t_\perp/a_0^2) \Delta
   f_\mathrm{pin} \, U.
  \label{eq:dFp}
\end{align}
Here, $\Delta f_\mathrm{pin}$ denotes the sum of jumps in the pinning force
when vortices jump into and snap out of the pinning trap created by a defect.
Furthermore, $t_\perp$ is the transverse length over which vortices passing by
the defect are trapped and $a_0^{-2} = B/\Phi_0$ is the vortex density
($\Phi_0 = hc/2 e$ denotes the flux unit). For a point-like defect, the
transverse trapping length is of the order of the vortex core size, $t_\perp
\sim \xi$.

The drop (\ref{eq:dFp}) in critical force appears whenever vortices start
moving to the left, i.e., when $U(X,t)$ decreases with increasing $t$.  In the
dynamical situation defined by the $ac$-response, we have to follow the
macroscopic displacement $U(X,t)$ in time and switch on the
restoring force
(\ref{eq:dFp}) when $U$ starts decreasing. Assume that vortices have reached
the displacement $U_0$ when they change direction of motion; the argument in
Eq.~(\ref{eq:dFp}) then has to be replaced by $U-U_0$, with the restoring
force smoothly growing from zero. In a fully dynamical situation, $U_0$ has to
be calculated self-consistently and is given by the maximal displacement
reached so far, $U_0(X,t) = \max_{t^\prime < t} U(X,t^\prime)$. The final
expression for the reduction in pinning force then is given by
\begin{align}\label{eq:dFp_final}
   \delta F_\mathrm{pin}(U)
   &= -\alpha_\mathrm{sp} \, (U-U_0) 
\end{align}
with
\begin{align}\label{eq:asp}
   \alpha_\mathrm{sp}
   &\approx n_p (t_\perp/a_0^2) \Delta f_\mathrm{pin}.
\end{align}
In the following, we solve the dynamical equation for $U(X,t)$, Eq.\
(\ref{eq:eomU}), with the restoring force $\delta F_\mathrm{pin}$, Eq.\
(\ref{eq:dFp_final}), derived from strong pinning theory.

\section{Critical State $ac$-Response}\label{sec:indyn}

Before entering the detailed discussion we give a short overview on the $ac$
dynamics.  Inserting the result (\ref{eq:dFp_final}) into Eq.\ (\ref{eq:eomU})
generates a complex vortex dynamics as flux enters the sample in a sequence of
diffusive pulses until the internal magnetic
field is raised to $B_0 + h_{ac}$---the
discussion of this initial dynamics is the central topic of the paper carried
out below. Once the asymptotic time domain has been reached, vortices exhibit
the typical oscillatory behavior within the pinning wells, but with respect to
the new critical state that has been shifted upward by $h_{ac}$, see Sec.\
\ref{sec:asymp} below. The $ac$ magnetic response then follows the standard
result with the Campbell length $\lambda_{\rm \scriptscriptstyle C}$
determined by the jump in pinning force $\Delta f_\mathrm{pin}$ through
$\alpha_\mathrm{sp}$, see Eq.\ (\ref{eq:asp}).

\subsection{Transient initialization regime}\label{sec:initial}

Our task is to solve the boundary-driven differential equation
\begin{align}\label{eq:de}
   - \partial_t U + D \partial_{\scriptscriptstyle X}^2 U 
   - \frac{\alpha_\mathrm{sp}}{\eta} (U-U_0) = 0
\end{align}
with the diffusion constant $D= B_0^2/4\pi\eta$, the max-field $U_0(X,t) =
\max_{t'<t} U(X,t')$, and the external drive $\partial_{\scriptscriptstyle X}
U(0,t) = -(h_{ac}/B_0) e^{-i\omega t}$. In a sample of finite thickness $d$
along $X$ vortices cannot move beyond the sample center, hence $U(d/2,t) = 0$
provides the second boundary condition.  We consider a situation where the
magnetic field $B_0$ changes little over the critical state profile, e.g., as it is the
case for a fully penetrated sample with thickness $d \ll L$, $L = c B_0/4\pi
j_c$ the asymptotic extension of the critical state profile, $B(X) = B_0 (L-X)/L$.
\begin{figure}[t]
\includegraphics[width=8.0cm]{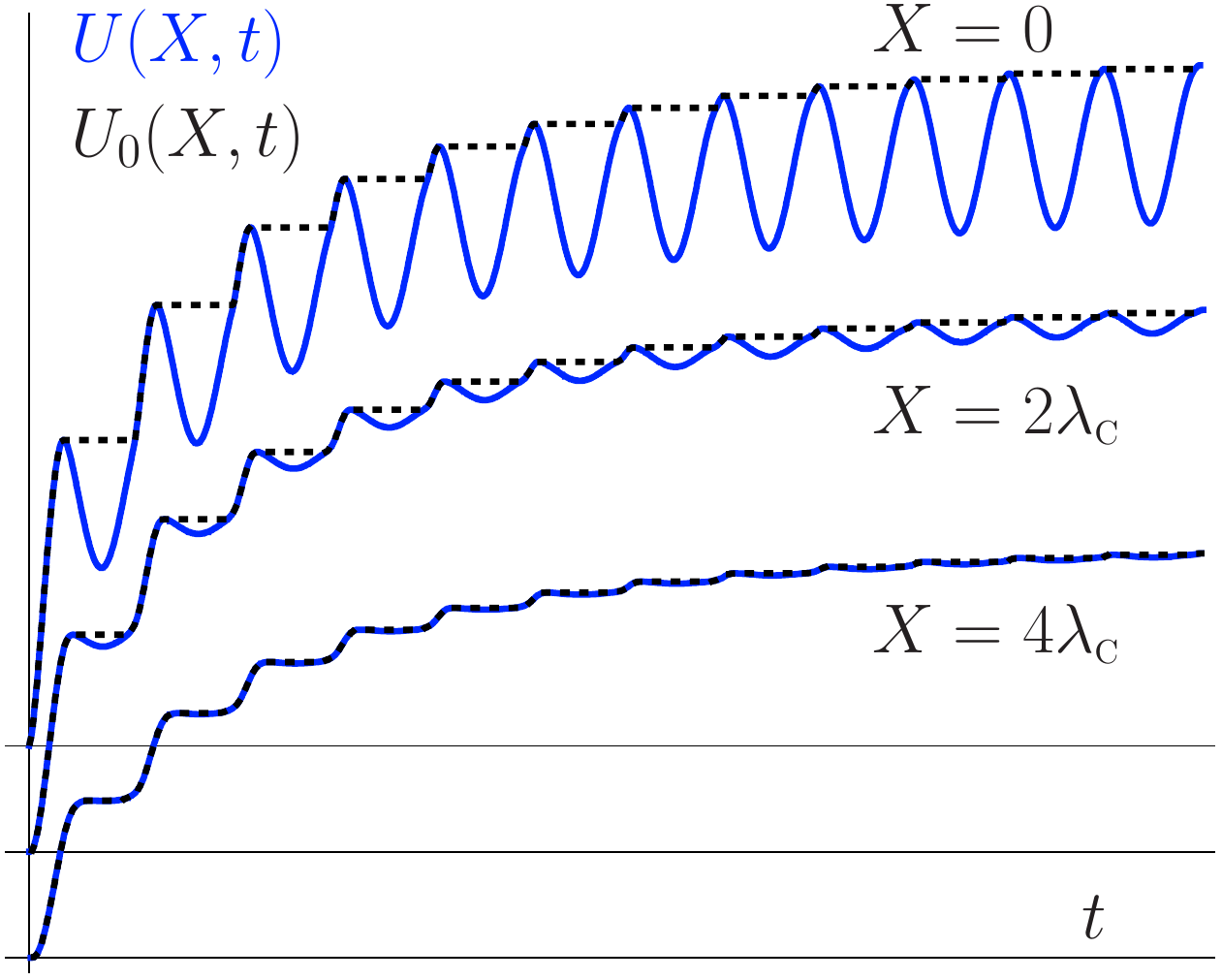}
\caption{Displacement $U$ (solid, blue) and max-field $U_0$ (black, dashed) as
a function of time $t$ at the positions $X=0$ (sample boundary),
$X=2\lambda_{\rm \scriptscriptstyle C}$, and $X=4\lambda_{\rm
\scriptscriptstyle C}$ (subsequent curves are shifted for better visibility).
The periodic decrease in $U$ describes relaxation of vortices in the pinning
wells. With the max-field $U_0$ remaining constant over these regions, a
reduction $\delta F_\mathrm{pin}$ in pinning force density below $F_c$ shows
up.}
\label{fig:U}
\end{figure}

Consider first the driven diffusion equation without the pinning term. The
(asymptotic) solution of this equation has the form $U(X,t) = (h_{ac}/B_0)
e^{-(1-i)(X/\delta)} e^{-i\omega t}$ with $\delta$ the skin depth involving
the flux-flow conductivity; the displacement $U(0,t)$ at the boundary
describes the net flux (per unit length along $Y$) $\phi(t) = B_0 U(0,t)$ that
has entered the sample up to time $t$---obviously flux periodically enters and
leaves the sample as well known from the skin effect.  

The pinning force term $-\alpha_\mathrm{sp}(U-U_0)$ breaks the symmetry
between vortices moving in and out of the sample. While flux entry proceeds as
before, flux exit is inhibited as the pinning force pushes the vortices into
the sample (due to the {\it reduction} of the critical force). As a result,
flux is periodically pumped into the sample until the total additional flux
reaches a value $\phi = h_{ac} d/2$ and the internal magnetic field is shifted
from $B_0$ to $B_0 + h_{ac}$.

\begin{figure}[b]
\includegraphics[width=8.0cm]{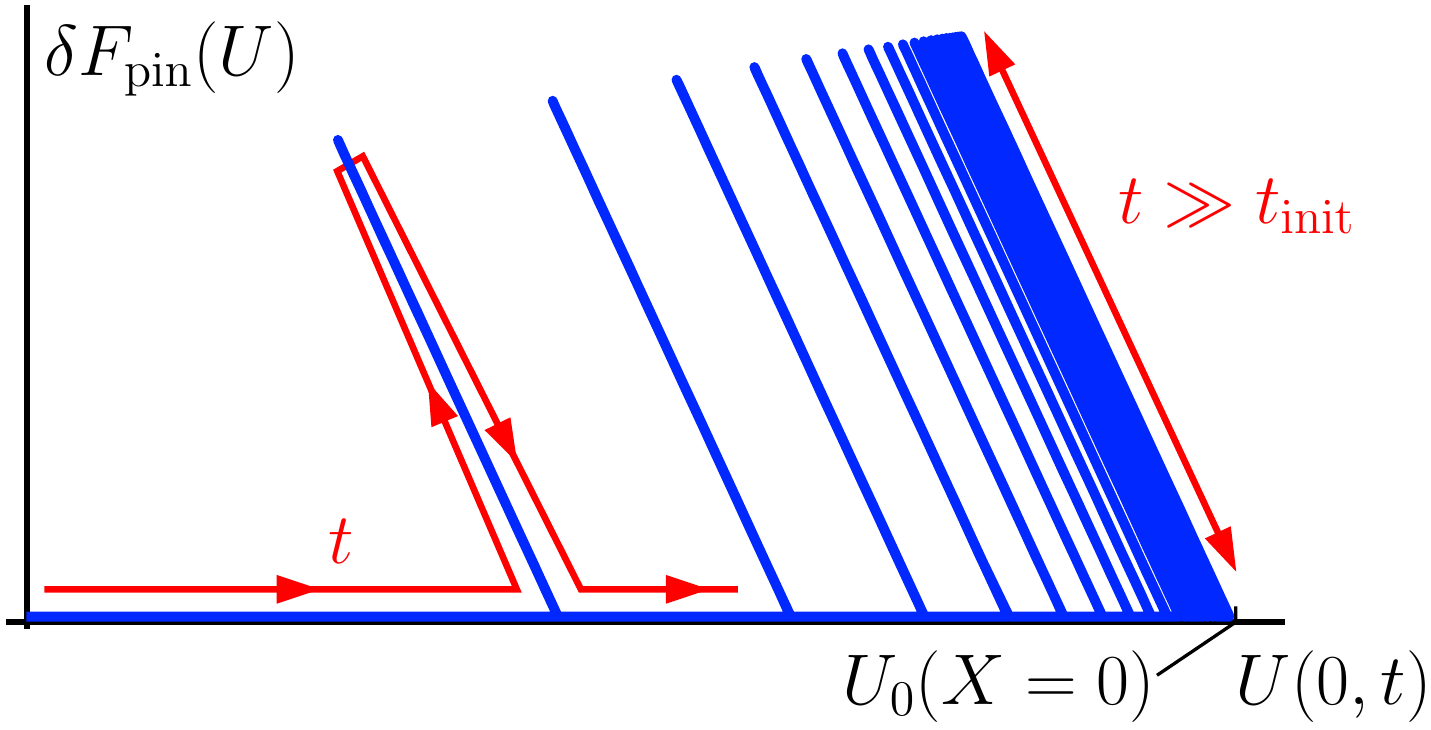}
\caption{Change in pinning force density $\delta F_\mathrm{pin}$ versus
displacement field $U$ at the sample surface $X = 0$. The time $t$ is a curve
parameter that evolves as indicated by the red line. Starting from $U=0$ and
increasing $U$, flux moves into the sample, the pinning force density is at
its maximum value $F_\mathrm{pin} = -F_c$, and $\delta F_\mathrm{pin} = 0$.
Deviations of $F_\mathrm{pin}$ from its critical value appear whenever $U$
decreases and vortices relax in their pinning potential.  At long times, the
Bean profile has been raised to the value $B_0 + h_{ac}$ and vortices
oscillate reversibly (see double-headed red arrow) within their potential
wells without additional flux entrance and vanishing regions with $\delta
F_\mathrm{pin} = 0$. The max-field $U_{0}(0,t)$ reaches its
asymptotic value $U_{0}(X = 0)$, see Eq.~\eqref{eq:U0_B}.}
\label{fig:F}
\end{figure}

\begin{figure}[tb]
\includegraphics[width=8.0cm]{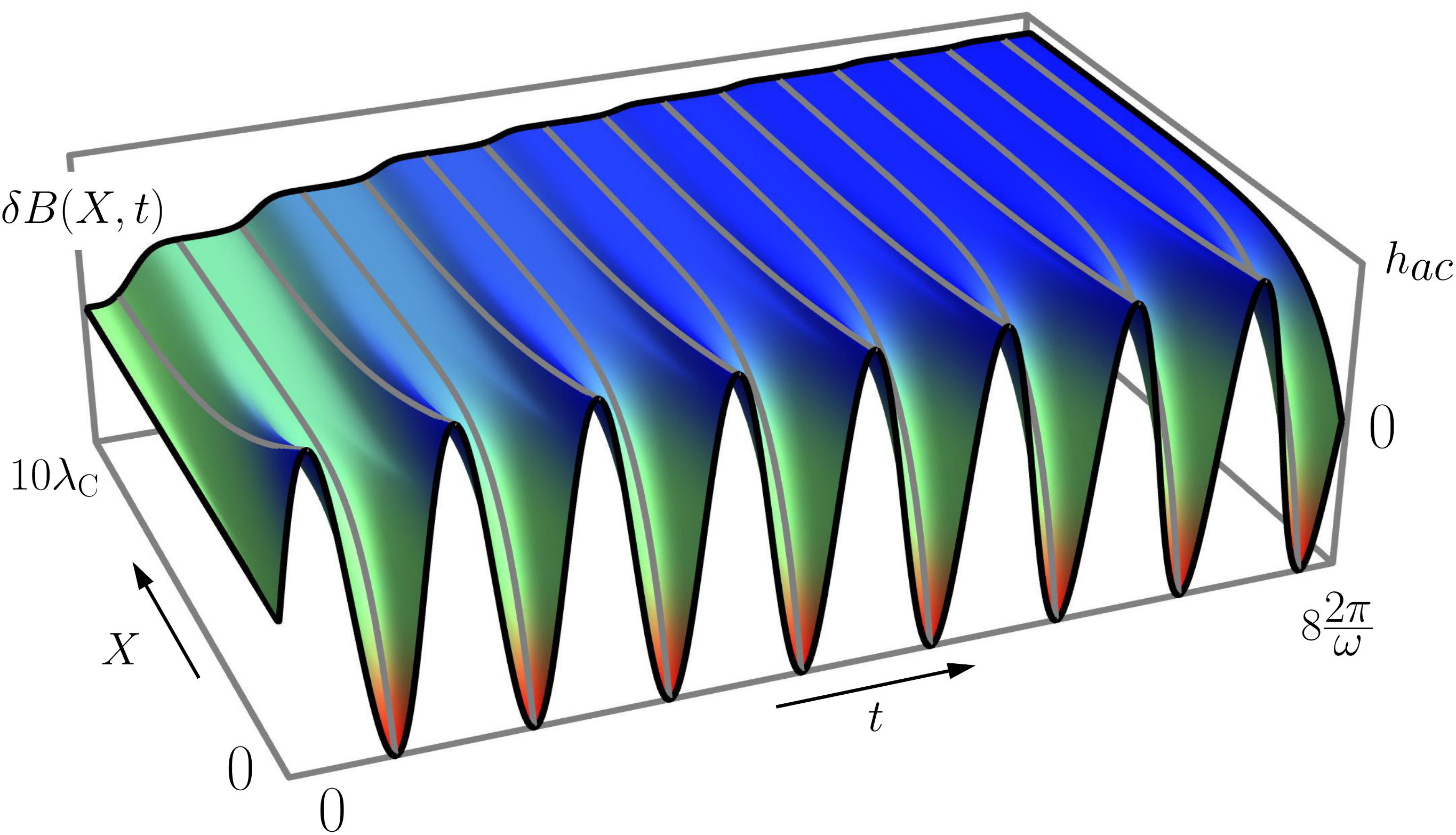}
\caption{Change in magnetic induction $\delta B(X,t)$ as a function of time
$t$ and distance $X$ into the sample. Every oscillation pumps flux into the
sample, pushing the critical state at $B_0$ ($\delta B = 0$, green) towards one at
$B_0 + h_{ac}$ ($\delta B = h_{ac}$, blue). The 8 cycles shown here push
almost all the flux into the sample that is required for its shift by
$h_{ac}$.}
\label{fig:B}
\end{figure}

\begin{figure}[tb]
\includegraphics[width=8.0cm]{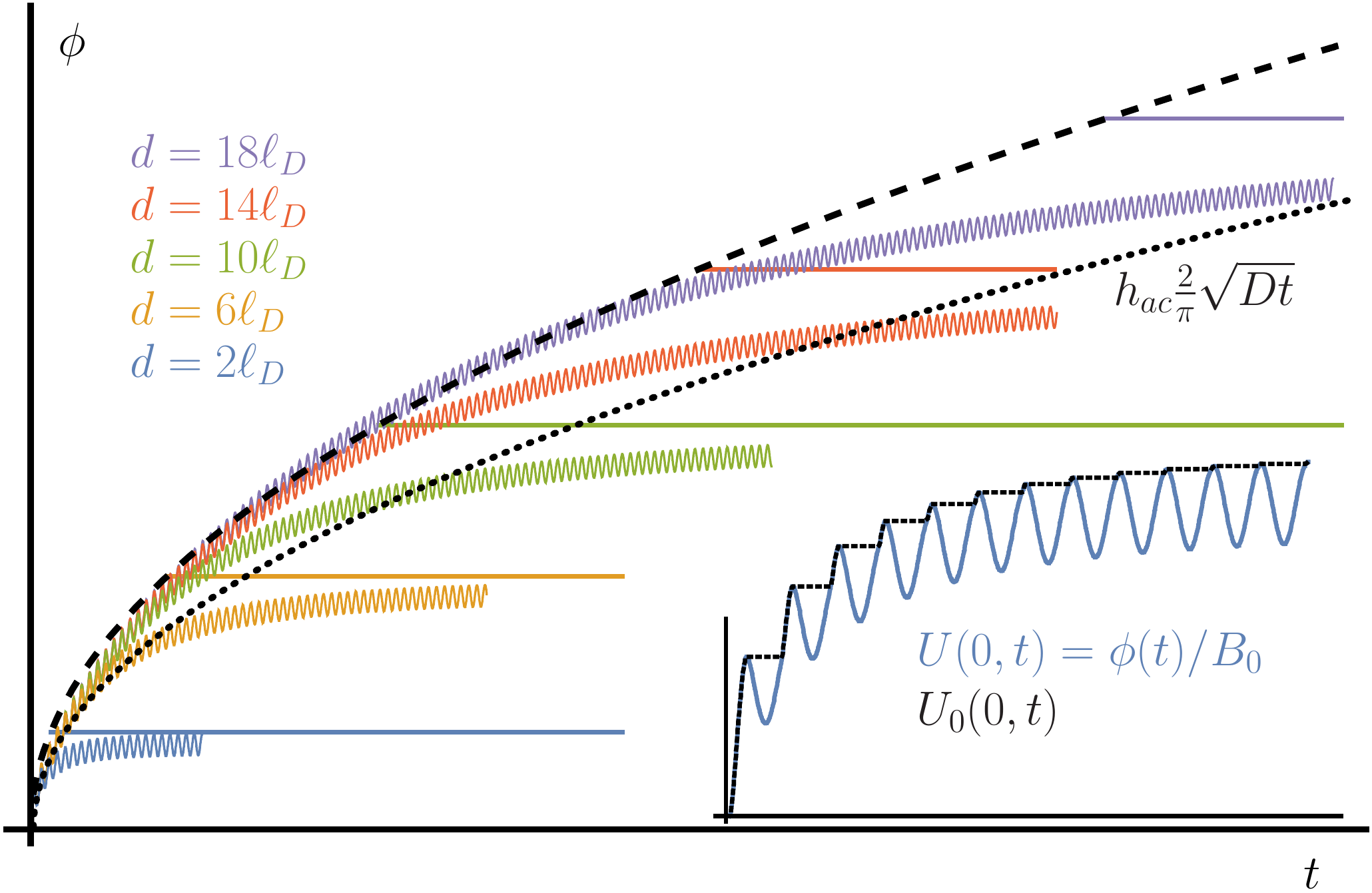}
\caption{Flux pumping $\phi(t) = B_0 U(0,t)$ as a function of time.  We
compare our numerical results (oscillating curves) for different sample sizes
$d$ with our analytical model (dotted line), see Eq.~\eqref{eq:phi_sol}. A better
quantitative agreement with the numerical data is obtained when scaling the
analytic result by $1.25$ (dashed line). The horizontal lines indicate for each
system the value of the penetrated flux $h_{ac} d/2$ after termination of the
initialization process. The inset shows an expanded view at small times of
both $U$ and $U_0$ at the surface.}
\label{fig:Phi-U}
\end{figure}

The differential equation for the displacement field $U(X,t)$ as well as the
max-field $U_0(X,t)$ can be found by numerical integration of Eq.\
(\ref{eq:de}). To this end, we bring Eq.\ \eqref{eq:de} into a dimensionless
form by measuring the time in units of the period $\tau = 2\pi / \omega$ and
lengths (and displacements) in units of the typical diffusion length $\ell_{D}
= \sqrt{D\tau}$ during the period $\tau$. The differential equation 
(\ref{eq:de}) then assumes the form
\begin{align}\label{eq:de-dimless}
   - \partial_t U
   + \partial_{\scriptscriptstyle X}^2 U 
   - \frac{\ell_{D}^{2}}{\lambda_{\mathrm{\scriptscriptstyle C}}^{2}} 
     (U-U_0) = 0,
\end{align}
with $t$, $X$, $U$, and $U_{0}$ the new dimensionless quantities and
$\lambda_{\rm \scriptscriptstyle C} = \sqrt{D \eta/\alpha_\mathrm{sp}} =
\sqrt{B_{0}^{2}/4\pi \alpha_{\mathrm{sp}}}$ the strong
pinning result for the Campbell length. The boundary condition at the surface
$X = 0$ now reads $\partial_{X}U(0,t) = - (h_{ac}/B_{0})e^{-i 2\pi t}$, while
the vanishing of the displacement at the sample center, $U(d/2\ell_{D},t) =
0$, inhibits vortices from further penetrating the sample. The relative
amplitude $(h_{ac}/B_{0})$ trivially scales the solution, while the ratios
$\lambda_{\mathrm{\scriptscriptstyle C}} / \ell_{D}$ and $d/2\ell_{D}$
determine the shape of the solution for a specific setup. The algorithm to
solve Eq.\ \eqref{eq:de-dimless} is based on the forward Euler routine on a
discrete mesh with spacing $\Delta X$ ($\Delta t$) in the spacial (time)
domain. The standard scheme
\begin{widetext}
\begin{align}\label{eq:forwardEuler}
   U(X, t + \Delta t)
       &= U(X, t)
          + \frac{\Delta t}{(\Delta X)^{2}}
            \big[U(X-\Delta X, t) - 2U(X, t) + U(X+\Delta X, t)\big]
          - \frac{\ell_{D}^{2} \Delta t}{\lambda_{\mathrm{\scriptscriptstyle C}}^{2}} 
            [U(X, t) - U_{0}(X, t)],
\end{align}
\end{widetext}
has to be supplemented with an update of the max-field
\begin{align}
 U_{0}(X, t + \Delta t) &= \max\{U(X, t + \Delta t), U_{0}(X, t)\}.
\end{align}
This additional rule does not allow for more advanced numerical approaches
such as the implicit Crank-Nicolson method. The physical constraint to resolve
the dynamics within the Campbell length $(\Delta X)^{2} \ll
(\lambda_{\mathrm{\scriptscriptstyle C}} / \ell_{D})^{2}$ sets a lower bound
for $\Delta X$. The requirement of good convergence of the numerical code
imposes a constraint on the mesh size $\Delta t$ via $\Delta t / (\Delta
X)^{2} \ll 1$ [these two constraints also guarantee that $\ell_{D}^{2} \Delta
t / \lambda_{\mathrm{\scriptscriptstyle C}}^{2} \ll 1$, see
Eq.~\eqref{eq:forwardEuler}]. A second constraint on the time mesh $\Delta t
\ll 1$, guaranteeing the smoothness of the external drive, is automatically
satisfied for the parameter range considered here. The results shown in Figs.
\ref{fig:U} -- \ref{fig:B} assume $\lambda_{\mathrm{\scriptscriptstyle
C}}/\ell_{D} = 1/\sqrt{20\pi} \approx 1/8$ and $d/2\ell_{D} = 1$. In Fig.
\ref{fig:Phi-U} different system sizes $d/2\ell_{D} =$ 1, 3, 5, 7, and 9 are
shown. 

The displacement fields $U(X,t)$ and $U_0(X,t)$ are shown in Fig.\ \ref{fig:U}
for several positions $X$, once at the boundary $X=0$ where $U$ is
proportional to the penetrated flux $\phi$ (per unit length along $Y$), as
well as a few penetration depths $\lambda_{\rm \scriptscriptstyle C}$
into the sample at $X = 2 \lambda_{\rm
\scriptscriptstyle C}$ and $X = 4 \lambda_{\rm \scriptscriptstyle C}$. Due to
the action of the pinning force term, the displacement $U$ is no longer
periodic: the increase in $U$ describing flux entry is interrupted by short
excursions with decreasing $U$ as vortices relax in their pinning potential.
The max-field $U_0$ ignores these depressions and the finite amplitude $U_0 -
U$ generates the decrease in pinning force $\delta F_\mathrm{pin}$. The latter
is shown in Fig.\ \ref{fig:F}: Regions of vanishing $\delta F_\mathrm{pin}$
where additional flux penetrates deeper into the sample are interrupted by
segments of finite force $\delta F_\mathrm{pin} > 0$ where vortices relax back
in their pinning potentials.  With increasing time, the additional flux shifts
the internal field upwards to $B_0 + h_{ac}$. For long times, vortices
evolve reversibly within their pinning wells, with no further flux entering
the sample. As a result, the change in pinning force density $\delta
F_\mathrm{pin}$ oscillates back and forth with maximal amplitude and no
intermediate regions with $\delta F_\mathrm{pin} = 0$ show up.

The physically most relevant and transparent result is the behavior of the
magnetic induction $B$ within the sample. In Fig.\ \ref{fig:B} we show the
evolution of $\delta B(X,t)$ with time $t$ over several penetration depths
into the sample along $X$. Additional flux is periodically pumped into the
sample, lifting the $dc$ value of $B$ with increasing time $t$ (change from
green to blue color). For better illustration we have chosen a geometry and
parameters such that the 8 cycles shown in the figure nearly suffice to reach
the asymptotic value $B_0 + h_{ac}$, see also the lowest curve in Fig.\
\ref{fig:Phi-U} for $d/2\ell{D} = 1$.

In the following, we build and analyze a simple model for the flux entry into
the sample; this will allow us to estimate the number of cycles needed to
reach the asymptotic state. We start with the first $ac$ cycle which pumps a
flux (per unit transverse length) $\phi_0 = f \, h_{ac} \ell_D$ into the
sample, where $\ell_D = \sqrt{D\tau}$ is the diffusion length during one
cycle, $\tau = 2\pi/\omega$, and $f<1$ is a numerical factor accounting for
the magnitude of penetrated flux per pulse (note that flux is pumped into the
sample only during the first half-cycle and the precise magnitude depends on
the detailed shape of the driving signal).  This flux pulse diffuses with time
$t$ and spreads over the distance $\sqrt{D t}$. At time $t$, the flux
pulse then reduces the $ac$-magnetic field amplitude at the sample boundary
$X=0$ by $\bar{h}_0(t) = \phi_0/\sqrt{D t}$.  Hence the next pulse
$\phi_1$ entering the sample is reduced, $\phi_1 = f \, [h_{ac}-\bar{h}_{0}
(\tau)] \ell_D$.  Iterating this process, the $i$-th flux pulse entering the
sample is given by
\begin{align}\label{eq:phi_i}
   \phi_i = f\, h_i \ell_D,
\end{align}
with the iteratively defined amplitude
\begin{align}\label{eq:h_i}
   h_i = h_{ac} - \sum_{m<i} \bar{h}_m(i\tau)
\end{align}
and 
\begin{align}\label{eq:bh_m}
   \bar{h}_m(t) = \phi_m/ \sqrt{D(t-m\tau)}.
\end{align}
Combining Eqs.\ \eqref{eq:phi_i}, \eqref{eq:h_i}, and \eqref{eq:bh_m} we then
have to solve the self-consistency equation for $h_i$,
\begin{align}\label{eq:h_is0}
   h_i = h_{ac} - f\sum_{m<i} \frac{h_m}{\sqrt{i-m}}.
\end{align}
Going over to continuous variables, we obtain the integral equation
\begin{align}\label{eq:h_is1}
   h(t) = h_{ac} - \frac{f}{\sqrt{\tau}}\int_{t_0}^t dt' \frac{h(t')}{\sqrt{t-t'}},
\end{align}
with the starting time $t_0$ to be determined together with $h(t)$. Inserting
the Ansatz $h(t) = \beta\, h_{ac}/\sqrt{t}$, we can carry out the integral and
arrive at the self-consistency condition
\begin{align}\label{eq:h_is2}
   \beta\, \frac{h_{ac}}{\sqrt{t}} =  
   h_{ac}\bigg[1- \frac{2 f \beta}{\sqrt{\tau}} 
   \arcsin\big(\sqrt{t'/t}\big)\Big|_{t_0}^t\bigg].
\end{align}
Choosing $\beta = \sqrt{\tau}/f \pi$ (such that the upper boundary $t$
provides a term $(2/\pi) \arcsin 1 = 1$) and setting $t_0 = \tau/2f^2$, we
obtain a consistent solution for large times\cite{footnote2} $t \gg t_{0}$ in
the form
\begin{align}\label{eq:h_sol}
   h(t) = \frac{h_{ac}}{f \pi}\sqrt{\frac{\tau}{t}}.
\end{align}
Finally, adding up the flux pulses $\phi_i$, we find that the flux penetrates
into the sample diffusively as
\begin{align}\label{eq:phi_sol}
   \phi(t) = \frac{2}{\pi}h_{ac} \ell_D \sqrt{t/\tau},
\end{align}
independent on $f$. The flux needed to push the internal
critical state from an initial field $B_0 d/2$ to a final field $(B_0+h_{ac})
d/2$ then involves the time
\begin{align}\label{eq:pen-time}
   t_{\mathrm{init}} = \frac{\pi^2}{16} \frac{d^2}{\ell_D^2} \tau
\end{align}
or $n = (\pi^2/16)\, d^2/\ell_D^2$ cycles. In Fig.\ \ref{fig:Phi-U} we compare
the result of our analytic model calculation with our numerical data.  For the
latter, we have computed the response of the vortex system for different
system sizes with the thickness $d/2$ of the critical state profile taking the values
$d/2\ell_{D} =$ 1, 3, 5, 7, and 9 in units of the diffusion length $\ell_{D}$.
Despite the simplicity of the model, the analytic result in Eq.\
\eqref{eq:phi_sol} is in good qualitative agreement with the numerical
results; for a quantitative agreement, the model result has to be multiplied
with a factor $\approx 1.25$.

Our comparison with recent experiments \cite{prozorov_13} refers to a
SrPd$_2$Ge$_2$ single crystal, a material that is isostructural to the Fe- and
Ni-pnictides, of typical size $d \sim 0.5~\mathrm{mm}$ in a field $H \sim
0.1~\mathrm{T}$ and subject to an $ac$ field $h_{ac} = 2~\mathrm{\mu T}$ at
frequencies $\nu = \omega/2\pi \approx 17 ~\mathrm{MHz}$.  With a normal state
resistivity \cite{sung_11} $\rho_n  \approx 20~\mu\Omega$cm and the upper
critical field \cite{prozorov_13} $H_{c2} \approx 0.5$ T, we obtain a
diffusion length $\ell_D \approx 40~\mu$m, hence $d/2\ell_D \sim 6$.  The
pinning strength is quantified through the Campbell penetration depth
measuring $\lambda_{\rm \scriptscriptstyle C} \approx 10~\mu$m, hence
$\lambda_{\rm \scriptscriptstyle C}/\ell_D \approx 1/4$. The elementary
displacement $U(0,\tau) \sim \ell_d (h_{ac}/H)$ is of the order of 8 {\AA} at
the beginning of the initial diffusion dominated regime and smaller,
$U(0,\tau) \sim \lambda_{\rm \scriptscriptstyle C} (h_{ac}/H) \approx 2$ {\AA}
in the asymptotic regime, far below the coherence length $\xi \approx 25$
{\AA}.  For such a sample and with the above experimental parameters we find
that the asymptotic state is reached within about one hundred cycles or
$t_{\mathrm{init}}\sim 10~\mu\mathrm{s}$, i.e., the transient initialization
regime is usually not interfering with the measurement of the Campbell length
$\lambda_{\rm \scriptscriptstyle C}$. Each of these cycles typically pumps a
fraction of a vortex into the sample, as $(\phi_0 d)/\Phi_0 \sim 1/5$.  To
complete the discussion, we briefly describe the result for the asymptotic
regime.

\subsection{Asymptotic periodic regime}\label{sec:asymp}

Once the asymptotic time domain has been reached, vortices exhibit the typical
oscillatory behavior near the surface, but with respect to the new critical
state that has been shifted upward by $h_{ac}$. Eq.~\eqref{eq:eomU} can be solved exactly for a half-infinite space and its solution
\begin{align}\label{eq:U_B}
   U(X,t) = U_0(X) - \lambda_{\rm\scriptscriptstyle C} (h_{ac}/B_0)
   e^{-X/\lambda_{\rm \scriptscriptstyle C}} [1-e^{-i \omega t}]
\end{align}
with the asymptotic max-field,
\begin{align}\label{eq:U0_B}
   U_0(X) = (\phi - h_{ac} X)/B_{0},
\end{align}
and the penetrated $dc$ flux $\phi$ provides a good approximation for a finite-size sample with $d/\lambda_{\mathrm{\scriptscriptstyle C}} \gg 1$. In Eq.~\eqref{eq:U_B}, the first term $U_{0}$ describes an overall shift of the critical state profile, $\delta B_0 = -B_0 \partial_{\scriptscriptstyle X} U_0 = h_{ac}$, and hence $\phi = h_{ac} d/2$. The second term in Eq.~\eqref{eq:U_B} captures
the oscillatory and decaying part of the magnetic field $- h_{ac} e^{-X/\lambda_{\rm
\scriptscriptstyle C}} [1-e^{-i \omega t}]$. Alternatively, the asymptotic
change of the magnetic field profile can be decomposed into an $ac$ component
$\delta B_{ac}(X,t) = h_{ac} e^{-X/\lambda_{\rm \scriptscriptstyle C}} e^{-i
\omega t}$ and an inhomogeneous $dc$ component $\delta B_{dc}(X) = h_{ac}[1-
e^{-X/\lambda_{\rm \scriptscriptstyle C}}]$. The conversion of the $ac$ drive
into a $dc$ signal is a unique feature of the critical state and may be
observed in an experiment. The Campbell penetration depth $\lambda_{\rm
\scriptscriptstyle C}$ is determined by $\alpha_\mathrm{sp}$, Eq.\
(\ref{eq:asp}), involving the jump in force $\Delta f_\mathrm{pin}$,
\begin{align}
   \lambda_{\rm \scriptscriptstyle C}^2 = 
   \frac{B_0^2}{4\pi \alpha_\mathrm{sp}} 
   \sim \frac{\lambda_{\mathrm{\scriptscriptstyle L}}^2}{n_p a_0 \xi^2 \kappa}
   \label{eq:lc}
\end{align}
with $\kappa$ the Labusch parameter, $\kappa \approx f_p a_0/\varepsilon_0
\xi$. Here, $f_p$ denotes the pinning force and $\varepsilon_{0} =
(\Phi_0/4\pi\lambda_{\mathrm{\scriptscriptstyle L}})^2$ is the vortex line energy. The small parameter $n_p
a_0 \xi^2 \kappa \ll 1$ defines the three-dimensional strong pinning regime
\cite{blatter_04}.

The jump in force $\Delta f_\mathrm{pin}$ in general depends on the state
preparation of the vortex system \cite{willa_15b}. Hence field-cooled and
zero-field cooled vortex states may exhibit different penetration depths and
hysteretic effects may show up. This better microscopic understanding of the
Campbell penetration depth may then allow to gain more detailed information on
the pinscape.

\subsection{Non-linear $ac$-response}
The above discussion has focused on Campbell penetration, i.e., the linear
response regime at small drive $h_{ac}$. As mentioned in the introduction,
this has to be distinguished from the Bean penetration at large amplitudes
$h_{ac}$.  The two regimes are separated by the conditions $\delta j \sim j_c$
or $h_{ac} \sim j_{c} \lambda_{\mathrm{\scriptscriptstyle C}}/c$, where the
vortex displacement $U \sim (h_{ac}/B_0) \lambda_{\rm\scriptscriptstyle C}$
matches the pinning length $x_+ \sim \kappa \xi$ and $\kappa >1$ is the strong
pinning parameter \cite{labusch_69}.  The Bean penetration at large amplitudes
exhibits similar behavior on a qualitative level.  E.g., using Bean's original
(quasistatic) approach, the $dc$ shift of the critical state profile by
$h_{ac}$ occurs during the first (quarter of the) $ac$ cycle; making use of a
more sophisticated model with a specific $I$-$V$ characteristics would result
in a penetration involving flux-pulses similar to those found above. At large
times, after the critical state has been shifted upwards by the $ac$ amplitude
$h_{ac}$, the Bean penetration reaches an oscillatory regime where the $ac$
drive generates a simple periodic dynamics on the penetration scale, see
Fig.~\ref{fig:CvsB} (right panel). On the other hand, the Bean penetration
differs from the Campbell penetration on a quantitative level: the
penetration depth $\ell_{\mathrm{\scriptscriptstyle B}} \sim c h_{ac}/ j_c$
within the Bean model scales with $h_{ac}$, resulting in a third harmonic $ac$
magnetic response \cite{tinkham_89}. The large displacements (compared to the
pinning length $x_{+}$) in the Bean regime do not resolve the internal
structure of the pinning centers and hence the $ac$ magnetic response is
independent of the state preparation.  Finally, the Bean penetration exhibits
generic hysteretic properties.

\section{Summary and conclusion}\label{sec:sc}

Using the results of strong pinning theory, we have studied the complete $ac$
magnetic penetration dynamics in the Campbell regime for a superconducting
sample prepared in a critical state. The field penetration proceeds in two
phases, a short (on the time scale of the experiment) transient initialization
regime where flux enters the sample in a sequence of pulses that shifts the
overall critical state by the $ac$ amplitude $h_{ac}$, followed by the
standard $ac$ dynamical regime where vortices merely move back and forth in
their pinning potentials with no additional net flux entering the sample.

The apparent divergence of the Campbell penetration depth
$\lambda_{\rm \scriptscriptstyle C}$ in the phenomenological description
\cite{prozorov_03} leaves its trace in the pulsed diffusive flux penetration
throughout the sample during the initialization regime. After long times,
the Campbell length $\lambda_{\mathrm{\scriptscriptstyle C}}$ is a (regular)
linear-response parameter containing valuable information on the pinscape
that can be extracted from its dependence on the state preparation
\cite{willa_15b}. Despite qualitative similarities between the
Campbell (at low $ac$ fields) and the Bean penetration (at higher $ac$
fields) the latter differs by i) its field-dependent penetration depth
$\ell_{\mathrm{\scriptscriptstyle B}} \sim c h_{ac}/ j_c$, ii) the generation
of a third harmonics $ac$ signal and iii) the insensitivity to the
pinscape structure. Finally, the rectified $dc$ signal $\delta B_{dc}(X)$
induced by the $ac$ (Campbell or Bean) drive is a signature that can be
probed in future experimental investigations.

\begin{acknowledgments}
We acknowledge financial support of the Fonds National Suisse through the NCCR
MaNEP.
\end{acknowledgments}


\begin{thebibliography}{99}

\bibitem{onnes_11} H.\ Kamerlingh Onnes, KNAW Proceedings {\bf 14 II}, 818 (1912).

\bibitem{meissner_33}  W.\ Meissner and R.\ Ochsenfeld, Naturwissenschaften {\bf 21},
787 (1933).

\bibitem{london_35} F.\ and H.\ London, Proc.\ Roy.\ Soc.\ (London) {\bf A149}, 71 (1935).

\bibitem{abrikosov_57} A.A.\ Abrikosov, 
Sov.\ Phys.\ JETP {\bf 5}, 1174 (1957).

\bibitem{shubnikov_37} L.V.\ Shubnikov, V.I.\ Khotkevich, Yu.D.\ Shepelev, and
Yu.N.\ Riabinin, Zh.\ Eksp.\ Teor.\ Fiz.\ {\bf 7}, 221 (1937).

\bibitem{campbell_69} A.M.\ Campbell, J.\ Phys.\ C {\bf 2}, 1492 (1969),
{\it ibid.} {\bf 4}, 3186 (1971).
%

\bibitem{bean_62} C.P.\ Bean, Phys.\ Rev.\ Lett.\ {\bf 8}, 250 (1962).

\bibitem{prozorov_03} R.\ Prozorov, R.W.\ Giannetta, N.\ Kameda, T.\ Tamegai,
J.A.\ Schlueter, and P.\ Fournier, Phys.\ Rev.\ B {\bf 67}, 184501 (2003).

\bibitem{prozorov_13} H.\ Kim, N.H.\ Sung, B.K.\ Cho, M.A.\ Tanatar, and R.\
Prozorov, Phys.\ Rev.\ B {\bf 87}, 094515 (2013).

\bibitem{willa_15b} R.\ Willa, V.B.\ Geshkenbein, R.\ Prozorov, and G.\ Blatter,
arXiv:1508.00757 [cond-mat.supr-con] (2015).

\bibitem{bardeen_65} J.\ Bardeen and M.J.\ Stephen,
Phys.\ Rev.\ {\bf 140}, 1197A (1965). 

\bibitem{labusch_69} R.\ Labusch, 
Cryst.\ Lattice Defects {\bf 1}, 1 (1969).

\bibitem{larkin_79} A.I.\ Larkin and Yu.N.\ Ovchinnikov, J.\ Low Temp.\ Phys.\
{\bf 34}, 409 (1979), A.I.\ Larkin and Yu.N.\ Ovchinnikov, in {\it
Nonequilibrium Superconductivity}, edited by D.N.\ Langenberg and A.I.\ Larkin
(Elsevier, Amsterdam, 1986), p.\ 493.

%
%
%
%
%
\bibitem{blatter_04} G.\ Blatter, V.B.\ Geshkenbein, and J.A.G.\ Koopmann,
Phys.\ Rev.\ Lett.\ {\bf 92}, 067009 (2004).

\bibitem{koshelev_11} A.E.\ Koshelev and A.B.\ Kolton,
Phys.\ Rev.\ B {\bf 84}, 104528 (2011).

\bibitem{thomann_12} A.U.\ Thomann, V.B.\ Geshkenbein, and G.\ Blatter,
Phys.\ Rev.\ Lett.\ {\bf 108}, 217001 (2012).

\bibitem{footnote2} At short times $\tau < t < t_{0}$ the right hand side of 
expression \eqref{eq:h_sol} should be cut off by $h_{ac}$.


\bibitem{sung_11} N.H.\ Sung, J.-S.\ Rhyee, and B.K.\ Cho,
Phys.\ Rev.\ B\ {\bf 83}, 094511 (2011).

%
%

\bibitem{tinkham_89} L.\ Ji, R.H.\ Sohn, G.C.\ Spalding, C.J.\ Lobb, and M.\ Tinkham,
Phys.\ Rev.\ B {\bf 40}, 10936 (1989).



\end{thebibliography}
\end{document}